\providecommand{\submissionmode}{1}
\newif\ifsubmission
\ifnum\submissionmode=0\relax
  \submissiontrue
\else
  \submissionfalse
\fi

\documentclass[orivec]{llncs}
\usepackage[T1]{fontenc}
\usepackage{microtype}
\usepackage{amsmath,amssymb,mathtools}

\usepackage{amsthm}
\usepackage{tikz,tikz-3dplot}
\usetikzlibrary{calc,arrows.meta}
\usepackage{xurl}
\usepackage{etoolbox}
\patchcmd{\UrlBreaks}{\do0\do1\do2\do3\do4\do5\do6\do7\do8\do9}{}{}{}
\usepackage[hidelinks]{hyperref}
\usepackage{orcidlink}
\renewcommand{\orcidID}[1]{\,\orcidlink{#1}}
\usepackage[nameinlink]{cleveref}
\usepackage{bbding} \usepackage[breakable]{tcolorbox}
\usepackage[boxed,commentsnumbered,noend,linesnumbered]{algorithm2e}

\crefname{claim}{claim}{claims}
\Crefname{claim}{Claim}{Claims}

\newenvironment{prob}[2][]
  {\begin{tcolorbox}[colback=white,colframe=black,
    boxrule=0.5pt,arc=0pt,left=6pt,right=6pt,top=4pt,bottom=4pt]
    \begin{problem}[#1]\label{#2}\begin{description}}
  {\end{description}\end{problem}\end{tcolorbox}}

\crefname{problem}{problem}{problems}
\Crefname{problem}{Problem}{Problems}

\newcounter{algo}

\newcommand{\Z}{\mathbb{Z}}
\newcommand{\Q}{\mathbb{Q}}
\newcommand{\R}{\mathbb{R}}
\newcommand{\Zge}{\Z_{\geq 0}}
\newcommand{\vect}[1]{\mathbf{#1}}
\newcommand{\one}{\vect{1}}
\newcommand{\poly}{\operatorname{poly}}

\newcommand{\supp}{\operatorname{supp}}
\newcommand{\calC}{\mathcal{C}}

\newcommand{\FPT}{\mathsf{FPT}}
\newcommand{\XP}{\mathsf{XP}}

\definecolor{latticeorange}{HTML}{D97706}
\definecolor{layerbottom}{HTML}{E3A098}
\definecolor{layerlower}{HTML}{D77B70}
\definecolor{layermiddle}{HTML}{C34D47}
\definecolor{layerupper}{HTML}{A82C35}
\definecolor{layertop}{HTML}{7F182B}
\definecolor{polyteal}{HTML}{309B85}
\definecolor{costpurple}{HTML}{7653A6}

\title{Integer Programming with Polyhedral Constraints on the Image}
\hypersetup{pdftitle={Integer Programming with Polyhedral Constraints on the Image}}
\ifsubmission
\author{Anonymous submission}
\authorrunning{Anonymous submission}
\institute{}
\hypersetup{pdfauthor={}}
\else
\author{Anubhav Dhar\inst{1,2}\orcidID{0009-0006-5922-8300}
        \and
        Tim Randolph\inst{3}\textsuperscript{(\Envelope)}\orcidID{0000-0003-4287-0680}
        \and
        Karol W\k{e}grzycki\inst{1}\orcidID{0000-0001-9746-5733}
}
\authorrunning{A. Dhar et al.}
\institute{Max Planck Institute for Informatics, Saarbr\"{u}cken, Germany\\
           \email{adhar@mpi-sws.org}, \email{kwegrzyc@mpi-inf.mpg.de}
           \and
           Saarland University, Saarbr\"{u}cken, Germany
           \and
           Harvey Mudd College, Claremont, CA, USA\\
           \email{trandolph@g.hmc.edu}}
\hypersetup{pdfauthor={Anubhav Dhar, Tim Randolph, Karol W\k{e}grzycki}}
\fi

\begin{document}
\maketitle

\begin{abstract}

We generalize integer programming to the setting in which the image $A \vect{x}$ is restricted to an explicitly given rational polyhedron $\mathcal{P}$. That is, we solve $\min\{\vect{c}^{\mathsf T}\vect{x}:A\vect{x}\in \mathcal{P},\ \vect{x}\in\Zge^n\}$, where $A\in\Z^{d\times n}$ and $\mathcal{P}$ is an explicitly given rational polyhedron. We present a randomized algorithm running in time $(d\Delta)^{O(d)}\poly(L)$ in this regime, where $\Delta \coloneqq  \|A\|_\infty$ and $L$ is the binary input length. This running time matches that of the best-known parameterized algorithms for integer programming, due to Eisenbrand and Weismantel and to Jansen and Rohwedder, and improves on the $O(d^2)$ exponent achieved by Dadush, L\'eonard, Rohwedder, and Verschae in a more general setting.

This result enables several applications, including exact algorithms for integer programming and unbounded knapsack in which the consumed resources $A\vect{x}$ must satisfy arbitrary additional linear constraints. We also obtain a randomized fixed-parameter algorithm for Unbounded Subset Sum parameterized by the doubling constant, resolving an open problem of Randolph and W\k{e}grzycki. For integer programs with $m$ constraints and bounded row doubling, we obtain an $f(\calC,m)\,n^{O(m)}\poly(L)$-time algorithm and show that the dependence on $m$ in the exponent of $n$ is necessary under $\mathsf{W[1]}\ne\FPT$.

\end{abstract}

\section{Introduction}\label{sec:intro}

Integer linear programming provides a framework for problems in
combinatorial optimization and the geometry of numbers~\cite{EW18}.
In standard form, it can be \mbox{written as}
\begin{equation}
    \label{eq:ilp}
    \min\{\vect{c}^{\mathsf T}\vect{x}:A\vect{x}=\vect{b},\ \vect{x}\in\Zge^n\},
\end{equation}
where $A\in\Z^{d\times n}$, $\vect{b}\in\Z^d$, and $\vect{c}\in\Q^n$.
Standard form can be generalized to include inequality constraints by adding slack variables. This problem is well known to be NP-hard~\cite{Karp72}, which motivates algorithms for restricted parameter regimes.

Classical algorithms exploit two complementary restrictions on the input: few
variables (small $n$) or few constraints with bounded coefficients (small $d$ and $\|A\|_\infty$). For example, Lenstra's
algorithm~\cite{Lenstra83} runs in polynomial time when the number of variables
is fixed (see also the recent breakthrough of Reis and Rothvoss~\cite{RR23}).
For integer programs with few constraints,
Papadimitriou~\cite{Papadimitriou81} gave a pseudopolynomial algorithm.  
Eisenbrand and Weismantel~\cite{EW18} used the Steinitz lemma to obtain a running time of
$(d\Delta)^{O(d)}\poly(L \cdot \|\vect{b}\|)$, where
$\Delta=\|A\|_\infty\geq1$ and $L$ is the binary input length (see also
\cite{JR19} for a better dependence on $\|\vect{b}\|$).  
Under the Almost-Exponential Time Hypothesis (see~\cite{Man26}), ILP requires
$2^{d^{1-o(1)}}\poly(L)$ time even when $A\in\{0,1\}^{d\times n}$ (see~\cite{KPW19}\footnote{Knop,
Pilipczuk, and Wrochna~\cite{KPW19} proved a tight ETH-based lower
bound excluding $2^{o(d\log d)}\poly(L)$-time algorithms, even for
$0/1$ constraint matrices. A recent preprint claims to refute
ETH~\cite{OpenAI26}, so we use the weaker Almost-ETH assumption here.}).

Given the success of existing parameterized algorithms in the small-$d$, small-$\Delta$ regime, it is natural to wonder whether we can achieve similar running times for more general problems. We are motivated by the following question:

\begin{center}
  \emph{
  \textbf{Question 1.} Can we generalize integer programming without compromising the parameterized running time?
  }
\end{center}

Variations on this question have already attracted significant attention in the literature. For example, Hunkenschr\"oder, Pokutta, and Weismantel~\cite{HPW23}  generalized the objective function to $g(A\vect{x})$, where $g$ is a sharp, convex function given by an oracle. Dadush, L\'eonard, Rohwedder, and
Verschae~\cite{DLRV23} consider the even more general objective
$g(A\vect{x})+\vect{c}^{\mathsf T}\vect{x}$ with arbitrary variable bounds.
However, for explicitly given $A$, these algorithms
run in $n^d(d\Delta)^{O(d^2)}\poly(L)$ time, significantly slower than the fastest known algorithms for integer linear programming~\cite{EW18,JR19}.\footnote{A recent preprint claims to solve integer programming with bounded variables in time $f(d)\Delta^{O(d)}\poly(L)$ for a computable function $f$~\cite{CMZ26}. However, to the best of our understanding, their techniques do not apply to the settings of Dadush et al.~\cite{DLRV23} or Hunkenschr\"oder~et~al.~\cite{HPW23}.}

\paragraph*{Our generalization.} We answer Question 1 in the affirmative when $A\vect{x}$ is constrained to an arbitrary, explicit rational polyhedron $\mathcal{P}$. To motivate our choice of setting, consider the following application. Suppose we would like to maximize the profit $\vect{p}^{\mathsf T}\vect{x}$ created by producing $n$ types of products, such as different car models. Each car model requires some quantity of each of $d$ resource types, such as steel, glass, and plastic. Each resource is subject to a global constraint. This situation is readily captured by an ordinary integer linear program.

However, when \emph{some constraints are linear functions of multiple resources}, their standard ILP formulation may have many constraints or large coefficients, losing the small-$d$, small-$\Delta$ running-time guarantee. For instance, restrictions on transportation and storage costs, payload limits, safe-operating thresholds, pollutant emissions caps, and percentage-based ethical sourcing laws might all apply to linear combinations of multiple resource types.

\begin{figure}[htbp]
\centering
\tdplotsetmaincoords{70}{100}
\begin{tikzpicture}[
  tdplot_main_coords,scale=.75,font=\small,
  line width=.45pt,line join=round,line cap=round,
  face/.style={fill=polyteal,fill opacity=.11,draw=none},
  polyedge/.style={draw=polyteal!75!black,line width=.55pt},
  sectionedge/.style={draw=polyteal!80!black,line width=.65pt}
]
  \coordinate (A) at (0,4.4,5.5);
  \coordinate (B) at (0,4.4,2.5);
  \coordinate (C) at (0,-.4,5.5);
  \coordinate (D) at (0,-.4,2.5);
  \coordinate (E) at (1.5,2,6.4);
  \coordinate (F) at (1.5,2,1.6);
  \coordinate (G) at (-1.5,2,6.4);
  \coordinate (H) at (-1.5,2,1.6);
  \coordinate (I) at (2.4,3.5,4);
  \coordinate (J) at (2.4,.5,4);
  \coordinate (K) at (-2.4,3.5,4);
  \coordinate (L) at (-2.4,.5,4);
  \coordinate (N) at (0,4.4,4);
  \coordinate (M) at (0,-.4,4);

\filldraw[fill=black!2,draw=black!20,line width=.3pt]
    (-4,-2,4)--(-4,6,4)--(4,6,4)--(4,-2,4)--cycle;
  \foreach \i in {-8,...,8}
    \foreach \j in {-4,...,12}
      \fill[black!38] ({\i/2},{\j/2},4) circle[radius=.24pt];
  \foreach \u/\v/\w in {A/I/B,F/I/B,F/I/J,F/D/J,C/D/J,
    C/E/J,I/E/J,I/E/A,G/E/A,G/E/C}
    \fill[face] (\u)--(\v)--(\w)--cycle;
  \fill[polyteal,fill opacity=.13] (N)--(I)--(J)--(M)--(L)--(K)--cycle;
  \draw[sectionedge] (N)--(I)--(J)--(M);
  \draw[polyteal!60!black,densely dashed] (M)--(L)--(K)--(N);
  \foreach \u/\v in {A/I,I/B,B/A,F/I,B/F,I/J,J/F,F/D,D/J,
    C/D,J/C,C/E,E/J,I/E,E/A,G/E,A/G,G/C}
    \draw[polyedge] (\u)--(\v);
  \foreach \i in {-2,0,2}
    \foreach \j in {-1,0,2,4,5}
      \fill[latticeorange] (\i,\j,4) circle[radius=1.6pt];
  \filldraw[fill=white,draw=latticeorange!70!black,line width=.7pt]
    (0,4,4) circle[radius=3.6pt];
  \fill[latticeorange] (0,4,4) circle[radius=1.8pt];
  \coordinate (costright) at (0,6.5,5.7);
  \begin{scope}[tdplot_screen_coords]
    \draw[-{Stealth[length=5pt]},costpurple,line width=.8pt]
      (costright)--++(-1.6,0)
      node[midway,above,text=costpurple] {cost};
  \end{scope}
\end{tikzpicture}
\caption{
Optimizing over images $\{A\vect{x} \; : \; \vect{x} \in \Z_{\geq 0}^n\}$ (orange points) within $\mathcal{P}$ (green icosahedron). The circled point minimizes cost over the feasible points.}
\label{fig:cone}
\end{figure}
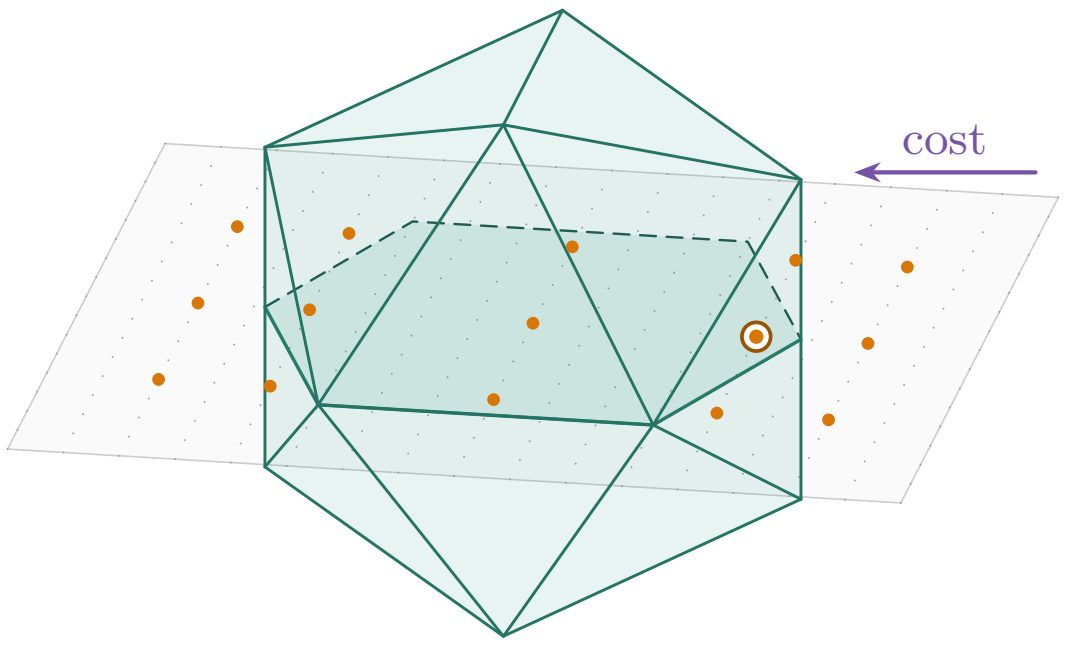

We can capture these situations by generalizing the resource constraint $A\vect{x} = \vect{b}$ to $A\vect{x} \in \mathcal{P}$,
where $\mathcal{P}$ is a polyhedron ($\mathcal{P} = \{\vect{y} \in \R^d \; : \; B \vect{y} \leq \vect{b}\}$
for a given $B \in \Q^{r\times d}$ and $\vect{b} \in \Q^r$).
Geometrically, this problem can be thought of as optimization over the integer
cone generated by the columns of $A$ with an additional polyhedral constraint
(see \Cref{fig:cone} for geometric intuition).

This formulation also naturally captures several other problems, including Unbounded Knapsack with multiple budgets (\Cref{cor:budget-knapsack}) and Unbounded Subset Sum parameterized by the doubling constant (\Cref{sec:small-doubling}).

\subsection{Summary of Contributions}

\begin{prob}[IP with Polyhedral Constraints (IP-PC)]{prob:main}
\item[Input:] $A\in\Z^{d\times n}$, $\vect{c}\in\Q^n$,
$B\in\Q^{r\times d}$, and $\vect{b}\in\Q^r$, defining the polyhedron
$\mathcal{P}=\{\vect{y}\in\R^d:B\vect{y}\leq\vect{b}\}$. 
\item[Output:] $\min\{\vect{c}^{\mathsf T}\vect{x}:
    A\vect{x}\in \mathcal{P},\ \vect{x}\in\Zge^n\}$.
\item[Parameters:] $d$ and $\Delta=\|A\|_\infty$.
\end{prob}
Here and below, an optimization problem asks for an optimal solution $\vect{x}$, or a report that none exists because the program is infeasible or unbounded; a feasibility problem asks for a feasible $\vect{x}$ or a report that none exists.

\begin{theorem}[IP with Polyhedral Constraints] \label{thm:main}
IP-PC can be solved in randomized time $(d\Delta)^{O(d)}\poly(L)$, with success probability at least $2/3$, where $L$ denotes the binary input length.
\end{theorem}

In other words, IP-PC is in randomized FPT with respect to the parameters $d$
(matrix height) and $\Delta$ (matrix entry size). The inequalities defining $\mathcal{P}$ affect the runtime only polynomially through the encoding length $L$. Note that this is asymptotically equivalent to the running time of the best existing algorithms~\cite{EW18,JR19}. The lower bound discussed above~\cite{Man26,KPW19} requires a $2^{d^{1-o(1)}}\poly(L)$ running time
under Almost-ETH, even for $\Delta=1$.

\begin{remark}[Comparison with \cite{DLRV23}]
IP-PC is a special case of the setting considered by Dadush et al.~\cite{DLRV23} in which $g$ is the indicator function of $\mathcal{P}$ (zero on $\mathcal{P}$ and $+\infty$ outside) and variables have no individual upper bounds. In this setting, merging equal columns in their algorithm gives $(d\Delta)^{O(d^2)}\poly(L)$ time. \Cref{thm:main} improves the exponent from $O(d^2)$ to $O(d)$.
\end{remark}

When $A$ is nonnegative, IP-PC becomes an unbounded multidimensional knapsack problem: each item consumes $d$ resources, and beyond simple capacity limits, arbitrary linear constraints may couple the amounts of different resources used. We obtain the same running time for this problem as a corollary of \Cref{thm:main}.

\begin{prob}[Knapsack with Polyhedral Constraints (K-PC)]{prob:knapsack-polyhedral}
\item[Input:] $A\in[0:\Delta]^{d\times n}$, $\vect{p}\in\Q^n$,
$B\in\Q^{r\times d}$, and $\vect{b}\in\Q^r$ with $d,\Delta\geq1$.
\item[Output:] $\max\{\vect{p}^{\mathsf T}\vect{x}:
     A\vect{x}\in\mathcal{P},\ \vect{x}\in\Zge^n\}$ for 
$\mathcal{P}=\{\vect{y}\in\R^d:B\vect{y}\leq\vect{b}\}$.
\item[Parameters:] $d$ and $\Delta$.
\end{prob}

\begin{corollary}[Knapsack with Polyhedral Constraints]\label{cor:budget-knapsack}
K-PC can be solved in randomized time $(d\Delta)^{O(d)} \poly(L)$ with success probability at least $2/3$, where $L$ denotes the binary input length.
\end{corollary}
This follows from \Cref{thm:main} by setting $\vect{c}=-\vect{p}$.

We next consider integer programming in settings with additive structure. Randolph and W\k{e}grzycki showed that when an instance of Unbounded Subset Sum has small \emph{doubling constant}, a condition that implies that the input integers have significant additive structure, the problem can be cast as an integer programming feasibility problem with $d = O_\calC(1)$ rows and entries of size $\Delta = O_\calC(n^{2/a})$ for a dimension $a = a(\calC)$, with the image restricted to a certain hyperplane \cite{randolph2024parameterized}.

\begin{prob}[$\calC$-Unbounded Subset Sum ($\calC$-USS)]{prob:uss}
\item[Input:] A set $Z = \{z_1,\ldots,z_n\} \subseteq \Zge$ and a target $t\in\Zge$.
\item[Output:] $\vect{x}\in\Zge^n$ satisfying
$\sum_{j=1}^n z_jx_j=t$.
\item[Parameter:] $\calC \coloneqq  |Z+Z|/|Z|$.\footnotemark
\end{prob}
\footnotetext{$Z+Z$ denotes the sumset $\{a + b \; : a, b \in Z\}$. The \emph{doubling constant} $\calC \coloneqq  |Z+Z|/|Z|$ captures the rate at which the set $Z$ ``expands'' under Minkowski addition and implies several equivalent properties, including that $Z$ can be contained in a small generalized arithmetic progression (see, e.g., \cite[Proposition~2.26]{tao2006additive}).}

Randolph and W\k{e}grzycki developed an $n^{O_\calC(1) \log \log \log (n)}$-time algorithm and left open whether $\calC$-USS can be solved in time $n^{O_\calC(1)}$ (that is, whether $\calC$-USS is in randomized $\XP$). We answer this question in the affirmative and go further by proving that Unbounded Subset Sum is in randomized $\FPT$ with respect~to~$\calC$.

\begin{corollary}
    \label{thm:uss}
    $\calC$-USS can be solved in randomized time $f(\calC) \cdot \poly(L)$ with one-sided error and success probability at least $2/3$, where $f$ is a computable function that depends only on $\calC$.
\end{corollary}

In fact, we can generalize our result for $\calC$-USS to any integer program in which each row of the constraint matrix has strong additive structure. Given a matrix $M \in \Z^{m \times n}$, let $\mathcal{M}_i$ denote the set of distinct entries in row $i$ and define the \emph{row doubling constant} $\calC \coloneqq \max_{i \in [m]} |\mathcal{M}_i+\mathcal{M}_i|/|\mathcal{M}_i|$. That is, $\calC$ is the maximum doubling constant of any row of $M$.

\begin{prob}[$\calC$-Row-Doubling ILP ($\calC$-ILP)]{prob:row-ilp}
\item[Input:] $M\in\Z^{m\times n}$ with row doubling at most $\calC$,
$\vect{b}\in\Z^m$, $\vect{c}\in\Q^n$, and rowwise relations
$\bowtie\,\in\{\leq,=,\geq\}^m$.
\item[Output:] $\min\{\vect{c}^{\mathsf T}\vect{x}:
    M\vect{x}\bowtie\vect{b},\ \vect{x}\in\Zge^n\}$.
\item[Parameters:] $\calC$ and $m$.
\end{prob}

\begin{corollary}
    \label{thm:row-ilp}\label{cor:rows}
    $\calC$-ILP can be solved in randomized time $f(\calC, m) \, n^{O(m)} \, \poly(L)$, where $f$ is a computable function that depends only on $\calC$ and $m$, and $L$ denotes the binary input length.
\end{corollary}

In contrast to \Cref{prob:main}, in $\calC$-ILP we do not treat $\|M\|_\infty$ as a small parameter. Instead, the strong additive structure implied by bounded row doubling effectively allows us to convert each row of $M$ into $O_\calC(1)$ rows with entries of size $O_\calC(n^{2/a})$ using the algorithmic ``Freiman compression'' of \cite{randolph2024parameterized}. The resulting (randomized) algorithm is fixed-parameter tractable with respect to $\calC$ and slice-wise polynomial with respect~to~$m$.

We also prove a lower bound for $\calC$-ILP (Appendix, \Cref{sec:clique}). The recently claimed refutation of the Exponential Time Hypothesis (ETH)~\cite{OpenAI26} would not settle whether $\mathsf{W[1]}=\FPT$. We exclude algorithms that take $f(m)\poly(L)$ time under $\mathsf{W[1]}\ne\FPT$, even for row doubling at most two (\Cref{thm:clique}), showing that the exponent of $n$ must depend on $m$.

\subsection{Our Techniques}

Our algorithm combines three ingredients. First, in \Cref{subsec:cand-solutions} we use a result of Rohwedder and W\k{e}grzycki \cite{RohwedderW25}, who build on Eisenbrand and Shmonin \cite{ES06} to conclude that every bounded, feasible instance of IP-PC has a \emph{lexicographically minimum} optimal solution $\vect{x}^*$ with support at most $O(d\log(d\Delta))$. Instead of trying all possible small supports, we then take advantage of the fact that our polyhedral constraint applies to a low-dimensional image: we can enumerate a set of candidate supports guaranteed to contain the support of the lexicographically minimum solution for \emph{every} image $A \vect{x} \in \Z^d$.

Every candidate support set gives us a low-dimensional sub-instance. Solving each sub-instance with a generic integer programming algorithm (e.g., \cite{RR23}) would be too slow, so we find a workaround. In \Cref{subsec:known-residues} we adapt an approximate integer programming algorithm due to Dadush, Eisenbrand, and Rothvoss \cite{DER25} to find a non-integral solution point in $\ell^{-1}\Z^n$ for a large integer $\ell$. This allows us to solve the problem by rescaling \emph{if} the residue class of the optimal solution modulo $\ell \Z^d$ is known.

Once again, we do not have time to try every residue, so we construct a representative family of residue classes covering the images of every residue in \Cref{subsec:res-enum}. Finally, in \Cref{subsec:main-alg}, we combine our ingredients to yield an algorithm for a threshold variant of IP-PC and reduce IP-PC to this problem to complete~the~proof of \Cref{thm:main}.

For the applications to sets with small doubling, we use the algorithmic
Freiman compression of Randolph and W\k{e}grzycki~\cite{randolph2024parameterized}. It rewrites each row with large coefficients as a constraint on the image over $O_\calC(1)$
coordinates whose entries are at most $h(\calC)\,n^{2/a}$.

\section{Preliminaries}\label{sec:prelim}

\paragraph{Notation.}
We write $[n]$ and $[a:b]$ for the integer sets $\{1,\ldots,n\}$ and  $\{a,a+1,\ldots,b\}$. 
Vectors are bold; their coordinates are not (e.g., $\vect{x} = (x_1, \ldots, x_n)$). For $\vect{x}\in\Zge^n$, we denote the \emph{support} by $\supp(\vect{x})=\{j:x_j>0\}$. For $S\subseteq[n]$, we let $\one_S\in\{0,1\}^n$ denote the incidence vector. $\vect{e}_j=\one_{\{j\}}$ denotes the $j$th unit vector. $I_n$ denotes the $n$-dimensional identity matrix.

We write $A_i$ to denote the $i$th column of the matrix $A$. Given a set $S \subseteq[n]$, we write $A_S$ to denote the submatrix of $A$ restricted to the column set indexed by $S$ and $\vect{c}_S$ to denote $\vect{c}$ restricted to indices $S$.

We write $B_2^{n}$ for the Euclidean unit ball.

\paragraph{Encoding.}
Throughout, $L$ denotes the total binary input length of the problem under
consideration, including dimensions, the rational data $A,\vect{c},B,\vect{b}$,
and additional objectives or thresholds. Our reductions 
increase this length only polynomially, which is absorbed
by $\poly(L)$. 
Since $n\leq L$, factors polynomial in $n$ are also absorbed into $\poly(L)$. For simplicity, bounds of the form $(d\Delta)^{O(d)}$ assume
$d,\Delta\ge1$ and should be read as $2^{O(d\log(2d\Delta))}$, so a constant
base such as $8^{d}$ is included even when $d\Delta=1$.

\paragraph{Randomization and one-sided error.} Our algorithms return only feasible points and succeed with probability at least $2/3$. For optimization, success means finding an optimum or correctly reporting infeasibility or unboundedness. Feasibility algorithms have one-sided error. The success probability can be amplified by repetition.

\paragraph{Representations and lexicographic minima.}

For an \emph{image} $\boldsymbol{\beta}\in\Z^d$, a \emph{representation} of
$\boldsymbol{\beta}$ is a vector $\vect{x}\in\Zge^n$ with $A\vect{x}=\boldsymbol{\beta}$. Because the feasibility of a vector $\vect{x}\in\Zge^n$ depends only on $A\vect{x}$, \emph{every} representation of a given image $\boldsymbol{\beta}$ is feasible if and only if $\boldsymbol{\beta} \in \mathcal{P}$.

We use lexicographic ordering to identify a single representation for each fixed image $\boldsymbol{\beta}$. For $n$-dimensional vectors $\vect{x}$ and $\vect{y}$, we write $\vect{x} \prec \vect{y}$ to indicate that $\vect{x}$ is lexicographically prior to $\vect{y}$. Given a cost vector $\vect{c}\in\Q^n$, we write $\vect{x} \prec_\vect{c} \vect{y}$ if $(\vect{c}^{\mathsf T}\vect{x},x_1,\ldots,x_n) \prec (\vect{c}^{\mathsf T}\vect{y},y_1,\ldots,y_n)$.

We use three properties of $\prec_{\vect{c}}$.
\begin{enumerate}
\item[(P1)] \emph{Total order.} If $\vect{x}\neq\vect{y}$, exactly one of
  $\vect{x}\prec_{\vect c}\vect{y}$ or $\vect{y}\prec_{\vect c}\vect{x}$ holds.
\item[(P2)] \emph{Translation invariance.} If $\vect{x}\prec_{\vect{c}}\vect{y}$,
  then $\vect{x}+\vect{z}\prec_{\vect{c}}\vect{y}+\vect{z}$ for every
  $\vect{z}\in\Z^n$. 
\item[(P3)] \emph{Existence of minima.} If $X\subseteq\Zge^n$ is nonempty
  with $\vect{c}^{\mathsf T}\vect x$ bounded below on $X$, then $X$ has a
  unique $\prec_{\vect c}$-minimum. 
\end{enumerate}

\section{Integer Programming with Polyhedral Constraints}

This section proves \Cref{thm:main}, following the outline in \Cref{sec:intro}.
\Cref{subsec:cand-solutions} constructs a small family of candidate supports,
\Cref{subsec:known-residues} solves a sub-instance when the residue of a
solution is known, \Cref{subsec:dup-cols} removes duplicate columns,
\Cref{subsec:res-enum} computes a small set of representative residues, and
\Cref{subsec:main-alg} combines these ingredients.

\subsection{Enumerating Solutions with Small Supports}
\label{subsec:cand-solutions}

Eisenbrand and Shmonin~\cite{ES06} proved that standard-form integer programs
with finite optima admit sparse optimal solutions. We use the following closely related statement due to Rohwedder and W\k{e}grzycki~\cite{RohwedderW25}, which specifies the existence of a \emph{lexicographically minimal} solution with a small support.

\begin{theorem}[{\cite[Theorem~3]{RohwedderW25}}]
    \label{thm:small-support}
    If the integer program \eqref{eq:ilp} is bounded and feasible, then the optimal solution $\vect{x}^*$ that is lexicographically minimum with respect to the cost vector $\vect{c}$ ($\prec_{\vect{c}}$) has support 
    \begin{equation}
        \label{eq:supp-bound}
        \|\vect{x}^*\|_0 \leq \Gamma \coloneqq \lceil 2(d+1)(\log (d+1) + \log \Delta + 2) \rceil.
    \end{equation}
\end{theorem}

Note that $\Gamma = O(d \log(d \Delta))$, so $\poly(\Gamma) = (d \log(d \Delta))^{O(1)} \leq (d \Delta)^{O(d)}$.

The following lemma allows us to quickly enumerate a small family of support sets guaranteed to include $\supp(\vect{x}^*)$. We follow the candidate-support argument of~\cite[Lemmas~7 and~8]{RohwedderW25} and defer the proof to \Cref{sec:prelim-apx}.

\begin{lemma}
    \label{lem:candidates}
    In time $(d\Delta)^{O(d)}\poly(L)$, we can construct a family of support sets $\mathcal{S}$ satisfying
    \begin{enumerate}
        \item $|\mathcal{S}| \leq (2\Gamma\Delta+1)^d$,
        \item $|S| \leq \Gamma$ for all $S \in \mathcal{S}$, and
        \item for every image $\boldsymbol{\beta} \in \Z^d$, $\mathcal{S}$ contains the support of the $\prec_{\vect{c}}$-minimum solution $\vect{x}^* \in \Z^n_{\geq 0}$ satisfying $A \vect{x}^* = \boldsymbol{\beta}$, if a $\prec_{\vect{c}}$-minimum solution exists.
    \end{enumerate}
\end{lemma}

In other words, $\mathcal{S}$ contains the support of at least one representation of \emph{every} possible image. This eliminates the need to know the image of an optimal solution ahead of time.

\subsection{Fast Integer Linear Programming with Known Residues}
\label{subsec:known-residues}

Every candidate support in the set $\mathcal{S}$ given by \Cref{lem:candidates} corresponds to a subproblem of dimension at most $\Gamma$. At this point, we could solve each subproblem in time $(\log \Gamma)^{O(\Gamma)}$ \cite{RR23}; however, substituting the bound on $\Gamma$ gives $(d\Delta)^{\omega(d)}$, exceeding our target runtime. Instead, we apply a corollary of an algorithm of Dadush, Eisenbrand, and Rothvoss for \emph{approximate} Integer Linear Programming. We use the following specialization of their result to $\Z^n$.
\begin{figure}[htbp]
\centering
\begin{tikzpicture}[font=\small,line cap=round,line join=round,>=Stealth,scale=0.8]
\foreach \panel/\offset in {0/-3.45,1/3.45} {
    \begin{scope}[xshift=\offset cm]
      \begin{scope}[x={(-.30494cm,-.61368cm)},y={(1.64531cm,-.11374cm)},z={(0cm,1.55258cm)}]
        \coordinate (a) at (-.65,-.65,-.65);
        \coordinate (b) at (.65,-.65,-.65);
        \coordinate (c) at (.65,.65,-.65);
        \coordinate (d) at (-.65,.65,-.65);
        \coordinate (e) at (-.65,-.65,.65);
        \coordinate (f) at (.65,-.65,.65);
        \coordinate (g) at (.65,.65,.65);
        \coordinate (h) at (-.65,.65,.65);
        \foreach \z in {-1,1}
          \foreach \i in {-1,0,1} {
            \draw[black!18,densely dotted] (\i,-1,\z)--(\i,1,\z);
            \draw[black!18,densely dotted] (-1,\i,\z)--(1,\i,\z);
          }
        \foreach \i in {-1,1}
          \foreach \j in {-1,1}
            \draw[black!18,densely dotted] (\i,\j,-1)--(\i,\j,1);
        \fill[polyteal!9] (a)--(b)--(c)--(d)--cycle;
        \fill[polyteal!17] (b)--(c)--(g)--(f)--cycle;
        \fill[polyteal!24] (c)--(d)--(h)--(g)--cycle;
        \fill[polyteal!12] (e)--(f)--(g)--(h)--cycle;
        \draw[polyteal!55!black,densely dashed] (a)--(b) (a)--(d) (a)--(e);
        \draw[polyteal!75!black] (b)--(c)--(d)--(h)--(e)--(f)--(b)
                                     (f)--(g)--(h) (c)--(g);
\ifnum\panel=0
          \def\gridstep{2}
        \else
          \def\gridstep{1}
        \fi
        \foreach \k/\shade in {-2/layerbottom,-1/layerlower,0/layermiddle,1/layerupper,2/layertop} {
          \pgfmathtruncatemacro{\showlayer}{mod(\k,\gridstep)}
          \ifnum\showlayer=0
            \foreach \i in {-2,...,2}
              \foreach \j in {-2,...,2} {
                \pgfmathtruncatemacro{\showpoint}{mod(abs(\i),\gridstep)+mod(abs(\j),\gridstep)}
                \ifnum\showpoint=0
                  \pgfmathtruncatemacro{\inside}{max(abs(\i),abs(\j),abs(\k))<2}
                  \ifnum\inside=1
                    \filldraw[fill=\shade,draw=white,line width=.4pt]
                      ({\i/2},{\j/2},{\k/2}) circle[radius=1.65pt];
                  \else
                    \draw[\shade,line width=.5pt]
                      ({\i/2},{\j/2},{\k/2}) circle[radius=1.05pt];
                  \fi
                \fi
              }
          \fi
        }
      \end{scope}
    \end{scope}
  }
\end{tikzpicture}
\caption{Illustration of~\Cref{thm:DER}. The green region depicts the same polytope $K$ in
both panels. Red markers are points in $\Z^n$ (left) or $\ell^{-1}\Z^n$ (right), with shades darkening from
bottom to top; filled dots lie inside $K$ and rings lie outside.}
\label{fig:DER}
\end{figure}
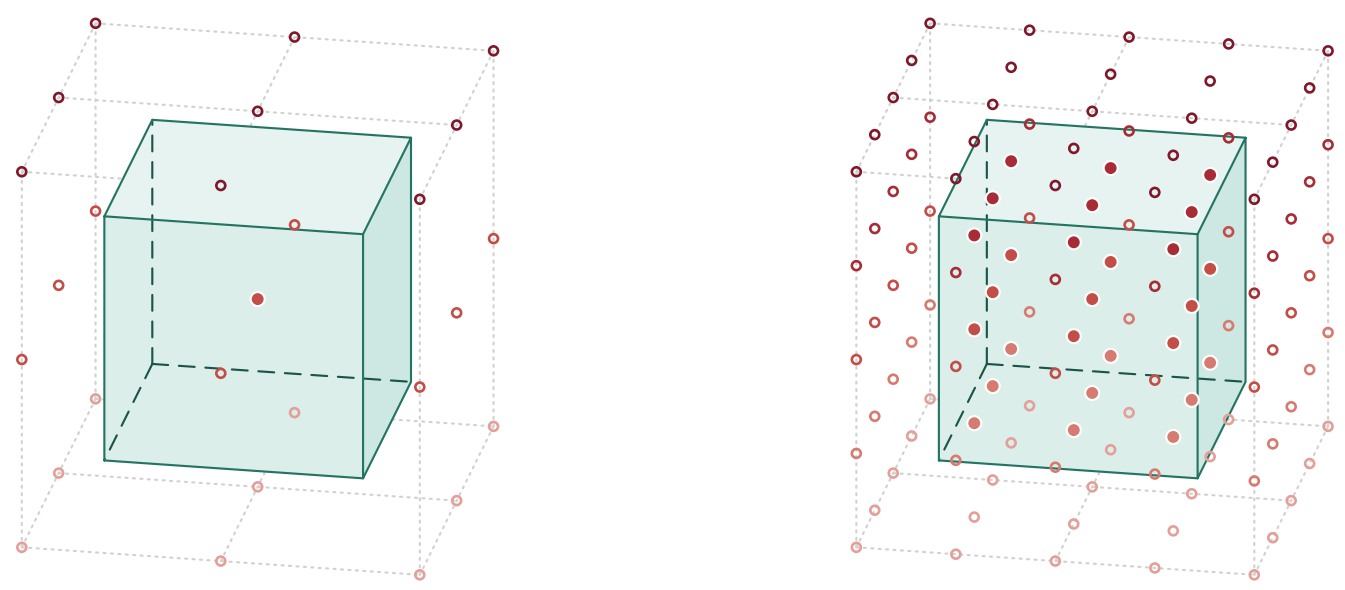

\begin{theorem}[{\cite[Theorem~16]{DER25}}]\label{thm:DER}
Given $\rho>0$, an integer $\ell\geq5(n+1)$, and a separation oracle for a closed convex set
$K\subseteq\rho B_2^{n}$, there exists a randomized algorithm that finds a point in $K\cap\ell^{-1}\Z^n$ or correctly reports
$K\cap\Z^n=\varnothing$ with high probability.
It runs in time $2^{O(n)}\poly(\log(2+\rho))$ times the cost of the separation oracle.
\end{theorem}

In other words, if the set $K$ contains a point in $\Z^n$, the algorithm finds a point in $\ell^{-1}\Z^n$ (\Cref{fig:DER}). We can use this algorithm to find an \emph{integer} point by guessing the residue of a target integer point modulo $\ell \Z^n$.

\begin{corollary}
    \label{cor:known-residues}
    Let $K \subseteq \R^n$ be an explicit rational polyhedron, $\ell = 5(n+1)$ be an integer, and $\vect{r} \in [0 : \ell - 1]^n$ be a residue modulo $\ell \Z^n$.

    There is a randomized $2^{O(n)} \poly(L)$-time algorithm that, if the set $K \cap (\vect{r} + \ell \Z^n)$ is non-empty, returns an integer point in $K$ with high probability.
\end{corollary}
\begin{proof}
    The idea behind the corollary is as follows: if we know the residue $\vect{r}$, then subtracting it from any point in $K \cap (\vect{r} + \ell \Z^n)$ yields a point in $\ell \Z^n$. Applying \Cref{thm:DER} with parameter $\ell$ then guarantees the existence of a point in $\ell^{-1}\Z^n$ that we can translate back to recover the original point.

    Fix an explicit rational polyhedron $K = \{ \vect{x} \; : \; G \vect{x} \leq \vect{g} \}$. Let $Q$ be the smaller polyhedron defined as follows:
    \[
        Q \coloneqq  \{ \vect{w} \in \R^n : \vect{r} + \ell \vect{w} \in K \}.
    \]
    $Q$ is a translated, scaled-down copy of $K$, in the sense that every integer point $\vect{x}^* \in K \cap (\vect{r} + \ell \Z^n)$ corresponds to the point $\vect{w}^*(\vect{x}^*) = (\vect{x}^* - \vect{r}) / \ell \in Q \cap \Z^n$.

    We still cannot apply \Cref{thm:DER} directly because $Q$ is not guaranteed to be bounded. Fortunately, a standard bound on the size of integer solutions\footnote{Papadimitriou's theorem is stated for systems $G'\vect{y}=\vect{g}'$, $\vect{y}\in\Z^{N}_{\geq 0}$ with integer data. Multiplying by a common denominator gives an integral system $G'\vect{w}\leq\vect{g}'$. Writing $\vect{w}=\vect{w}^+-\vect{w}^-$ with $\vect{w}^{\pm}\geq\vect{0}$ and adding a slack vector $\vect{t}\geq\vect{0}$ gives $G'\vect{w}^+-G'\vect{w}^-+\vect{t}=\vect{g}'$.} allows us to compute an integer $H$, of encoding length polynomial in that of $Q$, such that $Q$ contains an integer point if and only if $Q_H \coloneqq  Q \cap [-H, H]^n$ contains an~integer~point \cite{Papadimitriou81}.
    
    $Q_H$ is now a closed convex set contained in $[-H, H]^n \subseteq nH \cdot B_2^n$. We can create a separation oracle by checking the inequalities defining $Q_H$ in time polynomial in the encoding length. 
    
    Now we apply \Cref{thm:DER} with $\rho = nH$, $\ell = 5(n+1)$, and the closed convex set $Q_H$. If there exists an integer point $\vect{x}^* \in K \cap (\vect{r} + \ell \Z^n)$, then $\vect{w}^*(\vect{x}^*) \in Q \cap \Z^n$, and with high probability the algorithm returns a point $\vect{w} \in Q \cap \ell^{-1}\Z^n$. Check and return the point $\vect{x} \coloneqq  \vect{r} + \ell \vect{w}$, which is an integer point contained in $K$ by construction. By \Cref{thm:DER} the running time is $2^{O(n)} \cdot \poly(L)$.
\end{proof}

\subsection{Removing Duplicate Columns}\label{subsec:dup-cols}

Because duplicate columns of $A$ can be safely removed, we can bound $n$ by a function of $\Delta$ and $d$ without loss of generality.

\begin{lemma}
    \label{lem:dup-cols}
    Fix an instance $A \in \Z^{d \times n}$, $\vect{c} \in \Q^n$, $\mathcal{P}$ of IP-PC. We can compute an equivalent instance $A' \in \Z^{d \times n'}$, $\vect{c}'$, $\mathcal{P}$ satisfying $n' \leq (2\Delta + 1)^d$ in time $\poly(L)$, where $L$ denotes the binary encoding length of the input.
\end{lemma}
\begin{proof}
    As $\Delta \coloneqq \|A\|_\infty$, every column of $A$ is a vector in $[-\Delta : \Delta]^d$, so $A$ contains at most $(2\Delta + 1)^d$ \mbox{distinct columns}.

    Suppose $A$ contains two duplicate columns $A_i$, $A_j$ with $c_i \leq c_j$. We may assume without loss of generality that $x_j = 0$ in an optimal solution. If $x_j > 0$, we could set $x_i' \coloneqq x_i + x_j$ and $x_j' \coloneqq 0$, preserving the image of $\vect{x}$ without increasing the cost $\vect{c}^{\mathsf{T}} \vect{x}$. Thus from each class of duplicate columns, we keep only one of minimum cost and drop the others along with their entries of $\vect{c}$.
\end{proof}

\subsection{Enumerating Residues}
\label{subsec:res-enum}

Our known-residue algorithm (\Cref{cor:known-residues}) will allow us to achieve \Cref{thm:main} as long as we know the correct residue $\vect{r}$. However, even assuming $n \leq (2\Delta+1)^d$ (\Cref{lem:dup-cols}), there are still too many residues in $[0: \ell - 1]^n$ to enumerate and try them all within our target running time. We solve this problem by again using the fact that our polyhedral constraints apply only to a low-dimensional image. This allows us to construct a small set of representative residues that covers every residue image.

\begin{lemma}
    \label{lem:cand-residues}
    Fix an instance $A \in \Z^{d \times n}$, $\vect{c} \in \Q^n$, $\mathcal{P}$ of IP-PC and a support set $S \subseteq [n]$; define $n' \coloneqq |S|$, $A' \coloneqq A_S \in \Z^{d \times n'}$, $\vect{c}' \coloneqq \vect{c}_S$, and $\ell \coloneqq 5(n' + 1)$. 

    In time $|R| \cdot n' \ell \cdot \poly(L)$, we can compute a set of residues
    \[
        R \subseteq [0 : \ell - 1]^{n'}
    \]
    satisfying the following properties:
    \begin{enumerate}
        \item $|R| \leq (2n' (\ell - 1)\Delta + 1)^d$ and 
        \item for every residue $\vect{r}^* \in [0:\ell - 1]^{n'}$ there exists a representative $\vect{r} \in R$ with
        \[
            A' \vect{r} = A' \vect{r^*} \qquad \text{ and } \qquad \vect{c}'^{\mathsf{T}}\vect{r} \leq \vect{c}'^{\mathsf{T}}\vect{r}^*.
        \]
    \end{enumerate}
\end{lemma}
\begin{proof}
    We construct $R$ by dynamic programming on the column set of $A'$. For every $j \in [0:n']$, we will create a table that maps every feasible image $\boldsymbol{\beta} \in \Z^d$ to the cheapest vector $\vect{r} \in [0: \ell - 1]^j$ satisfying $\sum_{i \leq j} r_i A'_i = \boldsymbol{\beta}$.

    For $j = 0$, the only entry is $T_0[\vect{0}] = ()$. Given $T_{j-1}$, we can build $T_j$ as follows: for every entry $T_{j-1}[\boldsymbol{\beta}]$ and every $\rho \in [0 : \ell - 1]$, the candidate $T_{j-1}[\boldsymbol{\beta}] \circ \rho$ has image $\boldsymbol{\beta} + \rho A'_j$. Maintain the cheapest candidate for each feasible image, where $\circ$ denotes vector concatenation.

    Correctness follows from the fact that the $(j-1)$-dimensional prefix of $T_j[\boldsymbol{\beta}]$ has the image $\boldsymbol{\beta} - \rho^* A_j'$ for some $\rho^* \in [0:\ell - 1]$ and that this prefix must be cheapest for its image. 

    Every table cell is indexed by an image in $[-n'(\ell - 1) \Delta : n' (\ell - 1)\Delta]^d$ by construction, so $|R| \coloneqq |T_{n'}| \leq (2n'(\ell-1)\Delta + 1)^d$. Each table $T_{j-1}$ makes $|T_{j-1}|\cdot \ell \leq |R| \ell$ calculations to compute $T_j$, so the total process takes time $|R| \cdot n' \ell \cdot \poly(L)$.
\end{proof}

\subsection{Proof of \Cref{thm:main}}
\label{subsec:main-alg}

For simplicity, we give an algorithm for a threshold search variant of IP-PC. That is, given $A \in \Z^{d \times n}$, polyhedron $\mathcal{P}$, cost vector $\vect{c} \in \Q^n$ such that $\vect{c}^{\mathsf{T}}\vect{x}$ is bounded below on $\{\vect{x}\in\R^n_{\geq 0}:A\vect{x}\in\mathcal{P}\}$, and rational threshold $\tau \in \Q$, we seek a point $\vect{x} \in \Z^n_{\geq 0}$ such that $A\vect{x} \in \mathcal{P}$ and $\vect{c}^{\mathsf{T}}\vect{x} \leq \tau$. Given the correctness of Algorithm \ref{alg:ippc-threshold}, we can solve IP-PC via a standard reduction from optimization to threshold~search.

\begin{algorithm}[ht!]
    \DontPrintSemicolon
    \SetKwInput{KwFunction}{Function}
    \KwFunction{$\textsf{IP-PC-Threshold}(A \in \Z^{d \times n}, \vect{c}\in\Q^n, \mathcal{P}, \tau)$}
    \lIf{$\vect{0} \in \mathcal{P}$ and $0 \leq \tau$}{\Return $\vect{0}$}
    Compute the family $\mathcal{S}$ of candidate supports given by \Cref{lem:candidates}\;
    \ForEach{$S \in \mathcal{S}$}{
        Define $n' = |S|$, $A' \coloneqq A_S$, $\vect{c}' \coloneqq \vect{c}_S$, and $\ell = 5(n' + 1)$\;
        Compute the candidate residue set $R$ given by \Cref{lem:cand-residues}\;
        Define $K_S \coloneqq \{\vect{z} \in \R^{n'}_{\geq 0} : A' \vect{z} \in \mathcal{P}, \vect{c}'^{\mathsf{T}}\vect{z} \leq \tau\}$\;
        \ForEach{$\vect{r} \in R$}{
            Use \Cref{cor:known-residues} with residue $\vect{r}$ to search for an integer point $\vect{z}\in K_S$\;
            \lIf{a verified point is found}{\Return it, extended by zeros outside $S$}
        }
    }
    \Return failure\;
    \caption{Algorithm for IP-PC (threshold search variant).}
    \label{alg:ippc-threshold}
\end{algorithm}

\begin{proof}[Correctness of Algorithm \ref{alg:ippc-threshold}]
    We claim that Algorithm \ref{alg:ippc-threshold} returns a correct point with probability at least $2/3$ if and only if such a point exists.
    
    ($\Rightarrow$). $\vect{0}$ is returned only if it is feasible. Otherwise, any point $\vect{z}\in K_S$ found is integral by \Cref{cor:known-residues}, so its extension by zeros outside $S$ is a feasible output.

    ($\Leftarrow$). Suppose there exists a solution $\vect{x} \in \Z_{\geq 0}^n$ satisfying $A\vect{x} \in \mathcal{P}$ and $\vect{c}^{\mathsf{T}}\vect{x} \leq \tau$. Let $\vect{x}^*$ be the lexicographically minimal representation of $A \vect{x}$. If $\vect{x}^* = \vect{0}$, we return $\vect{0}$. Otherwise, by \Cref{lem:candidates}, $\vect{x}^*$ is supported on some set $S \in \mathcal{S}$.

    Then $A'\vect{x}^*_S = A\vect{x} \in \mathcal{P}$ and $\vect{c}'^{\mathsf T}\vect{x}^*_S = \vect{c}^{\mathsf T}\vect{x}^* \leq \vect{c}^{\mathsf T}\vect{x} \leq \tau$, so $\vect{x}^*_S \in K_S$. Write $\vect{x}^*_S$ as $\vect{x}^*_S = \vect{r}^* + \ell\vect{u}$, where $\vect{u}$ and $\vect{r}^*$ are integer vectors and $\vect{r}^*$ is the residue vector modulo $\ell \Z^{n'}$.

    By \Cref{lem:cand-residues}, there exists $\vect{r} \in R$ satisfying $A' \vect{r} = A'\vect{r}^*$ and $\vect{c}'^{\mathsf{T}}\vect{r} \leq \vect{c}'^{\mathsf{T}}\vect{r}^*$. Since $\vect{u}\geq\vect{0}$, the vector $\vect{z} \coloneqq \vect{r} + \ell\vect{u}$ is nonnegative and satisfies $A'\vect{z}=A'\vect{x}^*_S$ and $\vect{c}'^{\mathsf{T}}\vect{z}\leq\vect{c}'^{\mathsf{T}}\vect{x}^*_S\leq\tau$. Thus $\vect{z}\in K_S \cap (\vect{r} + \ell \Z^{n'})$, so when the inner loop reaches this residue, \Cref{cor:known-residues} returns an integer point of $K_S$ with high probability.
\end{proof}

\begin{proof}[Runtime of Algorithm \ref{alg:ippc-threshold}]
    Line 1 is a single feasibility check that takes time $\poly(L)$. By \Cref{lem:candidates}, Line 2 takes time $(d\Delta)^{O(d)} \poly(L)$ and the outer loop in Line 3 runs at most $(2\Gamma\Delta+1)^d = (d\Delta)^{O(d)}$ times.
    
    For each candidate support $S$, Line 5 generates a set $R$ of size $(2n'(\ell - 1)\Delta + 1)^d$. Substituting $\ell = 5(n'+1) = O(\Gamma)$ yields $|R| = (d\Delta)^{O(d)}$, and generating $R$ takes time $(d\Delta)^{O(d)} \poly(L)$, by \Cref{lem:cand-residues}. For every element $\vect{r} \in R$, \Cref{cor:known-residues} runs in time $2^{O(n')} \poly(L) = 2^{O(\Gamma)} \poly(L) = 2^{O(d \log(d\Delta))} \poly(L) = (d\Delta)^{O(d)} \poly(L)$. Multiplying all three terms together yields a total running time of $(d\Delta)^{O(d)} \poly(L)$ for the loop in Line 3 of Algorithm \ref{alg:ippc-threshold}.
\end{proof}

\begin{proof}[Proof of \Cref{thm:main}]
    Apply \Cref{lem:dup-cols} to ensure $A$ has at most $(2\Delta + 1)^d$ \mbox{distinct columns}.

    Next, run Algorithm \ref{alg:ippc-threshold} with $\vect{c} = \vect{0}$ and $\tau = 0$ to obtain a feasible point $\vect{x}^0$, or report infeasibility. Then solve the linear relaxation $\min\{\vect{c}^{\mathsf{T}}\vect{x} \; : \; A\vect{x} \in \mathcal{P},\ \vect{x} \in \R^n_{\geq 0}\}$. If it is unbounded, there exists an integer vector $\vect{u} \geq \vect{0}$ with $B A \vect{u} \leq \vect{0}$ and $\vect{c}^{\mathsf{T}}\vect{u} < 0$. This implies $\vect{x}^0 + t \vect{u}$ is feasible for all $t \in \Z_{\geq 0}$ and its cost approaches $-\infty$ as $t \rightarrow \infty$.

    Finally, if the linear relaxation is bounded, scale $\vect{c}$ to be integral and use Algorithm \ref{alg:ippc-threshold} to binary search for an integer solution in the range between the LP optimum and $\vect{c}^{\mathsf{T}}\vect{x}^0$.
    Repeating each of the $\poly(L)$ calls, including initial feasibility, $O(\log L)$ times independently gives overall success probability at least $2/3$ by a union bound.
\end{proof}

\section{Further Applications}\label{sec:applications}\label{sec:small-doubling}

In this section, we prove~\cref{thm:row-ilp,thm:uss} by 
combining~\Cref{thm:main} with the constructive Freiman reduction
of~\cite{randolph2024parameterized}, which gives algorithms for integer programs with
small doubling. 

Recall that for a nonempty finite set $Z\subseteq\Z^r$, its doubling constant is
$\calC(Z)=|Z+Z|/|Z|$, where
$Z+Z=\{\vect{u}+\vect{v}:\vect{u},\vect{v}\in Z\}$.
A matrix has \emph{row doubling} at most $\calC$ if the set of distinct
coefficients in each row has doubling at most $\calC$.

We use the following form of the constructive Freiman reduction.

\begin{lemma}[{\cite[Theorem~5, Observation~6, and Lemma~10]{randolph2024parameterized}}]\label{lem:freiman}
For a nonempty set $Z=\{z_1,\ldots,z_n\}$ of positive integers with doubling
at most $\calC$, a Monte Carlo algorithm running in time $f(\calC)\poly(L)$
constructs, with probability at least $2/3$, a matrix $A\in\Zge^{a\times n}$
and a vector $\vect{s}\in\Z^a$ such that
\[
    \|A\|_\infty\leq h(\calC)n^{2/a} \text{ and }
    z_j=\vect{s}^{\mathsf T}A_j \text{ for all } j\in[n],
\]
where $a=a(\calC)\geq1$ and $h=h(\calC)\geq1$ are computable functions.
\end{lemma}

The dimension $a$ can be chosen uniformly by padding the generalized
arithmetic progression in the cited construction before splitting its
long coordinates into digits. For a singleton, use the first unit vector
as its coordinate and put the input integer in the first coordinate of
$\vect{s}$.

\begin{proof}[Proof of~\Cref{thm:row-ilp}]
Let $M\vect{x}\bowtie\vect{b}$ be the input constraints.
Translate row $i$ by $\tau_i=1-\min_j M_{ij}$, making its coefficients
positive without changing its doubling constant. Apply~\Cref{lem:freiman}
to the distinct translated coefficients in each row, giving repeated
coefficients the same coordinates. Write $\vect{s}_i$ for the step vector
of row $i$, and put $a=a(\calC)$. For each row $i$ and variable $j$, we
check that $M_{ij}+\tau_i$ equals $\vect{s}_i^{\mathsf T}$ times the
coordinate vector of $M_{ij}+\tau_i$, and that every coordinate vector has
entries at most $h(\calC)n^{2/a}$; if any check fails, we report failure,
so the reduction never produces an incorrect instance. Stack the coordinate vectors of each
variable across all rows and append a $1$. This gives
$A\in\Zge^{d\times n}$ with $d=ma+1$ and
$\Delta\leq h(\calC)n^{2/a}$.

Write $\vect{y}^{(i)}$ for the block of $a$ coordinates associated with
row $i$, and set
\[
    \mathcal{P}=\{\vect{y}\in\R^d:
    \vect{s}_i^{\mathsf T}\vect{y}^{(i)}-\tau_i y_d
    \bowtie_i b_i\text{ for all }i\in[m]\}.
\]
(Each equality constraint can be written as two inequalities, so $\mathcal{P}$ has the form required by IP-PC.)
For $\vect{y}=A\vect{x}$, the last coordinate is $\sum_jx_j$, so
$\vect{s}_i^{\mathsf T}\vect{y}^{(i)}-\tau_i y_d=(M\vect{x})_i$.
Thus $M\vect{x}\bowtie\vect{b}$ if and only if $A\vect{x}\in \mathcal{P}$,
and the objective is unchanged.
The reduction has encoding length $f(\calC,m)\poly(L)$, so
\Cref{thm:main} gives
\[
    (d\Delta)^{O(d)}f(\calC,m)\poly(L)
       =f(\calC,m)n^{O(m)}\poly(L). \qedhere
\]
\end{proof}

\begin{proof}[Proof of~\Cref{thm:uss}]
Apply~\Cref{thm:row-ilp} with $m = 1$ to the single equation
$\sum_jz_jx_j=t$ with zero costs. Checking the returned vector gives
one-sided error.
\end{proof}

\begin{credits}
\subsubsection{\ackname}
\ifsubmission\else
Anubhav Dhar and Karol W\k{e}grzycki are supported by the Deutsche Forschungsgemeinschaft (DFG, German Research Foundation) -- project number 559177164.
\fi

The arguments in this work were developed by the authors in collaboration with ChatGPT 6.1 and Claude Opus 5.5. In particular, an LLM brought to our attention that \Cref{cor:known-residues} follows by a simple argument from the result of Dadush, Eisenbrand, and Rothvoss~\cite{DER25}. The manuscript was written by the authors, who referenced AI-generated text during the writing process. The authors take responsibility for the correctness of the work.
The authors permit reviewers to upload this submission to generative AI services.

\subsubsection{\discintname}
The authors have no competing interests to declare that are relevant to the content of this article.
\end{credits}

\bibliographystyle{splncs04}
\bibliography{main}

\appendix
\section{Proof of \Cref{lem:candidates}}\label{sec:prelim-apx}

\begin{proof}
    Fix an integer linear program as in \eqref{eq:ilp} and consider a two-dimensional table $T$ indexed by 
    \begin{enumerate}
        \item a support size parameter $s \in [0:\Gamma]$ and
        \item an image $\boldsymbol{\beta} \in [-\Gamma \Delta : \Gamma \Delta]^d$.
    \end{enumerate}

    For each pair $s, \boldsymbol{\beta}$, define $T[s, \boldsymbol{\beta}]$ to be the lexicographically minimum solution vector with $\ell_1$-norm at most $s$: 
    \begin{equation}
        \label{eq:DP}
        T[s, \boldsymbol{\beta}] \coloneqq  \min_{\prec_{\vect{c}}}\{\vect{y} \in \Z^n_{\geq 0} \; | \; A \vect{y} = \boldsymbol{\beta}, \|\vect{y}\|_1 \leq s\},
    \end{equation}
    with $T[s, \boldsymbol{\beta}]$ left undefined if no feasible vector exists.

    Fix $s$ and $\boldsymbol{\beta}$ such that $T[s, \boldsymbol{\beta}]$ has a well-defined, nonzero value. There must exist an index $j \in [n]$ at which $T[s, \boldsymbol{\beta}]$ is positive. We claim that
    \[
        T[s, \boldsymbol{\beta}] - \vect{e}_j = T[s-1, \boldsymbol{\beta} - A_j],
    \]
    where $\vect{e}_j$ denotes the $j$th unit vector. To see this, observe that $A(T[s, \boldsymbol{\beta}] - \vect{e}_j) = \boldsymbol{\beta} - A_j$ and $\|T[s, \boldsymbol{\beta}] - \vect{e}_j\|_1 \leq s - 1$. Moreover, because the relation $\prec_{\vect{c}}$ is preserved by addition, $(T[s, \boldsymbol{\beta}] - \vect{e}_j)$ must be the lexicographically minimum vector with these properties.

    Set $T[0, \vect{0}] = \vect{0}$ and leave all other entries with $s=0$ undefined. For $s \in [1:\Gamma]$ and every $\boldsymbol{\beta}$, including $\boldsymbol{\beta}=\vect{0}$, use the recurrence relation
    \begin{equation}
        \label{eq:DP-recurrence}
            T[s,\boldsymbol{\beta}]=\min_{\prec_{\vect{c}}}\bigl(
                   \{T[s-1,\boldsymbol{\beta}]\}\cup\{T[s-1,\boldsymbol{\beta}-A_j]+\vect{e}_j:j\in[n]\}
               \bigr),
    \end{equation}
    ignoring undefined entries and lookups outside the bounds of the table. We return the family $\mathcal{S}$ containing the support set of every defined table entry $T[\Gamma, \boldsymbol{\beta}]$. This implies $|\mathcal{S}| \leq (2\Gamma\Delta+1)^d$, as claimed.
    
    Because the table has at most $(\Gamma+1)(2\Gamma \Delta + 1)^d$ cells and filling each cell takes $n$ table lookups, the dynamic program takes time
    \[
        n(\Gamma+1)(2\Gamma \Delta + 1)^d \poly(L) = (d \Delta)^{O(d)} \poly(L),
    \]
    as claimed.

    It remains to show that $\mathcal{S}$ contains the support of the lexicographically minimum solution $\vect{x}^* \in \Z^n_{\geq 0}$ for every image $\boldsymbol{\beta} \in \Z^d$. Fix $\vect{x}^*$, note that $\|\vect{x}^*\|_0 \leq \Gamma$ by \eqref{eq:supp-bound}, and let $\vect{z}=\one_{\supp(\vect{x}^*)}$. First observe that $\|\vect{z}\|_1 \leq \Gamma$, so $A \vect{z} \in [-\Gamma \Delta : \Gamma \Delta]^d$; in other words, $T[\Gamma, A\vect{z}]$ is a valid cell in the table.

    We claim $\supp(\vect{z}) \in \mathcal{S}$. To see this, let $\vect{y} \coloneqq  T[\Gamma, A\vect{z}]$ and assume for contradiction that $\vect{z} \neq \vect{y}$. Because $A \vect{y} = A \vect{z}$, we have $A(\vect{x}^* - \vect{z} + \vect{y}) = A \vect{x}^* = \boldsymbol{\beta}$. Moreover, because $\vect{x}^* - \vect{z} \geq \vect{0}$, $\vect{x}^* - \vect{z} + \vect{y} \geq \vect{0}$; in other words, $\vect{x}^* - \vect{z} + \vect{y}$ is feasible for $\boldsymbol{\beta}$. However, since $\vect{y} \prec_{\vect{c}} \vect{z}$ by assumption, we have $\vect{x}^* - \vect{z} + \vect{y} \prec_{\vect{c}} \vect{x}^*$, contradicting the minimality of $\vect{x}^*$. Therefore $\vect{z} = T[\Gamma, A\vect{z}]$ and $\supp(\vect{z}) \in \mathcal{S}$.
\end{proof}

\section{Clique Lower Bound}\label{sec:clique}

For row doubling, bounded additive structure alone cannot remove the
number of constraints from the exponent of $n$. The following result
complements~\Cref{cor:rows} and holds even for feasibility. It does not
follow from the lower bound of~\cite{KPW19}, which uses $0/1$ constraint
matrices: for these, integer programming is fixed-parameter tractable
in $m$~\cite{EW18}. Our reduction instead uses coefficients of size
polynomial in $N$, and it keeps the row doubling below two by completing
each coefficient set to an interval.

\begin{theorem}\label{thm:clique}
Under $\mathsf{W[1]}\ne\FPT$, integer programming with $m$ equations, non-negative integer variables, and row doubling at most two
has no $f(m)\poly(L)$-time algorithm for any computable $f$.
\end{theorem}

We use the standard consequence of $\mathsf{W[1]}\ne\FPT$
that deciding whether an $N$-vertex graph contains a clique of size $k$
has no $f(k)N^{O(1)}$-time algorithm for any computable $f$.
This lower bound transfers to \textsc{Multicolored Clique} by making
one copy of every vertex in each color
class and joining copies whose original vertices are adjacent. The resulting
graph has $kN$ vertices, with the same parameter $k$.

We reduce from \textsc{Multicolored Clique}. Let $G$ have nonempty color
classes $V_1,\ldots,V_k$, $N$ vertices, and $k\geq2$.
Delete edges within color classes and number the vertices by $[N]$.

Introduce a nonnegative integer variable $x_v$ for each vertex and $y_e$
for each edge. There are $g=k+\binom{k}{2}$ groups: the $k$ vertex classes
and the edge sets between color pairs. List the latter as
$E_1,\ldots,E_{g-k}$, including empty sets. Impose
\begin{equation}\label{eq:selection}
\begin{aligned}
    \sum_{v\in V(G)}x_v+\sum_{e\in E(G)}y_e &=g,\\
    \sum_{i=1}^k(g+1)^{i-1}\sum_{v\in V_i}x_v
    +\sum_{j=1}^{g-k}(g+1)^{k+j-1}\sum_{e\in E_j}y_e
        &=\sum_{j=1}^g(g+1)^{j-1}.
\end{aligned}
\end{equation}
For an edge $e$ meeting $V_i$, write $V_i(e)$ for its endpoint in $V_i$.
For each color $i$, add the two constraints
\begin{align}
    (k-1)\sum_{v\in V_i}(N-v)x_v
      +\sum_{\substack{e\in E(G)\\e\cap V_i\neq\emptyset}}V_i(e)y_e
      &=(k-1)N,\label{eq:first-moment}\\
    (k-1)\sum_{v\in V_i}(N^2-v^2)x_v
      +\sum_{\substack{e\in E(G)\\e\cap V_i\neq\emptyset}}V_i(e)^2y_e
      &=(k-1)N^2.\label{eq:second-moment}
\end{align}
To complete the coefficient sets to intervals, introduce a nonnegative variable
$z_i$ for each integer $i$ from zero to the largest coefficient in these
$2k+2$ equations. Add $\sum_i i z_i$ to the left-hand side of every equation
and impose
\begin{equation}\label{eq:completion}
    \sum_i z_i=0.
\end{equation}
Every old row now has an interval as its coefficient set, and the new row has
coefficient set $\{0,1\}$, so all rows have doubling below two.

\begin{proof}[Proof of~\Cref{thm:clique}]
A multicolored clique gives a feasible solution by setting its vertex and edge
variables to one and all other variables to zero. Conversely,
\eqref{eq:completion} forces every $z_i=0$, and~\eqref{eq:selection} bounds
each group sum by $g$, so uniqueness of base-$(g+1)$ expansions forces one
selection per group. For each color, \eqref{eq:first-moment} and
\eqref{eq:second-moment} give the $k-1$ selected endpoint labels zero variance
and mean equal to the selected vertex's label. Thus every selected edge joins
the selected vertices, which form a multicolored clique.

The constructed instance has $m=2k+3$, $n=O_k(N^2)$, and encoding length
and construction time $O_k(N^3)$. A hypothetical
$f(m)\poly(L)$-time algorithm would therefore solve
\textsc{Multicolored Clique} in $N^{O(1)}$ time up to a computable factor
depending only on $k$, contradicting clique hardness.
\end{proof}

\begin{remark}
The same reduction would turn an $f(m)n^{o(m)}\poly(L)$-time algorithm for
this problem into an $f(k)N^{o(k)}$-time algorithm for $k$-Clique. Such an
algorithm remains unknown, even after the recently claimed refutation of
ETH~\cite{OpenAI26}.
\end{remark}

\end{document}